\documentclass[3p,times,authoryear]{elsarticle}

\usepackage{ecrc,amsmath}

\volume{00}

\firstpage{1}

\journalname{Planetary and Space Science}

\runauth{Q.-Z. Ye}

\jid{pss}

\jnltitlelogo{Planetary and Space Science}

\usepackage{amssymb,url,aas_macros,lineno}
\usepackage[figuresright]{rotating}

\begin{document}

\begin{frontmatter}



\dochead{}

\title{Meteor showers from active asteroids and dormant comets in near-Earth space: a review}


\author{Quan-Zhi Ye}

\address{Division of Physics, Mathematics and Astronomy, California Institute of Technology, Pasadena, CA 91125, U.S.A.\\
Infrared Processing and Analysis Center, California Institute of Technology, Pasadena, CA 91125, U.S.A.}

\begin{abstract}
Small bodies in the solar system are conventionally classified into asteroids and comets. However, it is recently found that a small number of objects can exhibit properties of both asteroids and comets. Some are more consistent with asteroids despite episodic ejections and are labeled as ``active asteroids'', while some might be aging comets with depleting volatiles. Ejecta produced by active asteroids and/or dormant comets are potentially detectable as meteor showers at the Earth if they are in Earth-crossing orbits, allowing us to retrieve information about the historic activities of these objects. Meteor showers from small bodies with low and/or intermittent activities are usually weak, making shower confirmation and parent association challenging. We show that statistical tests are useful for identifying likely parent-shower pairs. Comprehensive analyses of physical and dynamical properties of meteor showers can lead to deepen understanding on the history of their parents. Meteor outbursts can trace to recent episodic ejections from the parents, and ``orphan'' showers may point to historic disintegration events. The flourish of NEO and meteor surveys during the past decade has produced a number of high-confidence parent-shower associations, most have not been studied in detail. More work is needed to understand the formation and evolution of these parent-shower pairs.
\end{abstract}


\end{frontmatter}

%\linenumbers

\section{Introduction}

The term \textit{small solar system bodies} includes most natural bodies in the solar system that are less than a few hundred kilometers in size such as asteroids and comets. Traditionally, the word \textit{asteroid} refers to the rocky bodies that orbit the Sun between between the orbits of Mars and Jupiter and appear star-like, while the word \textit{comet} refers to the icy bodies in planet-crossing orbits that exhibit fuzzy atmosphere (\textit{coma}) and sometimes a tail as they approach the Sun. As it has been recently noticed, the boundary between asteroids and comets is blurry: some asteroidal objects can suddenly exhibit comet-like activities while objects in comet-like orbit appear asteroidal. It has been suggested that the activity from asteroids can be driven by sublimation of subterranean ice, impacts by a secondary body, as well as rotational or thermal excitation \citep{Jewitt2015}, while inactive objects in cometary orbits are thought to be ex-comets that have depleted their volatile ice \citep{Weissman2002}.

Active small body releases dust or \textit{meteoroids} into interplanetary space, forming a \textit{meteoroid~stream} along the orbit. For small bodies in Earth-crossing orbits, the ejected meteoroids may find their way to the Earth and produce \textit{meteor showers} as they plunge into Earth's atmosphere. Observation of a meteor shower provides information about the past activity of its parent. Even if a small body is observationally inactive at the moment, detection of associated meteor activity can provide evidence of recent dust production of this body. This is particularly useful for the study of objects with intermittent activities and/or have recently ceased to be active.

Attempt to link meteor showers to observationally inactive bodies goes back to \citet{Whipple1938}. Readers may refer to \citet{Jenniskens2008} for a historical account on this topic. More recently, the operation of dedicated near-Earth object (NEO) surveys has led to the discovery of a number of dual-designated objects that were initially identified as asteroids but were later found to exhibit cometary activity\footnote{The International Astronomical Union (IAU)'s Minor Planet Center defines dual-designated objects as objects concurrently holds \textit{permanent} designation of both comets and asteroids, \url{http://www.minorplanetcenter.net/iau/lists/DualStatus.html}. Here we use a more relaxed definition of dual-designation: any comets that hold asteroidal provisional designation are considered as dual-designated objects.}. Most of these objects are \textit{bona~fide} comets that are simply difficult to resolve at large distances due to low activity.

As of 2017 November, the IAU Meteor Data Center or MDC \citep{Jopek2011,Jopek2014,Jopek2017} lists 703 meteor showers, among which 112 are considered as ``established'' while most others are considered in the working list. The established showers are of high confidence and therefore we only focus on these showers, though we note that promotions from working list to established showers happen once every 3 years (during the IAU General Assembly, with the next one in 2018), therefore our list might miss a few newly established showers. According to MDC, a total of 15 established showers have been proposed to associate to asteroids, in addition another 2 have been linked to dual-designated objects. We tabulate these linkages in Table~\ref{tbl:shr} as listed on MDC, with the exception of the new linkage of $\alpha$ Capricornids --- 2017 MB$_1$ which is not being listed as of this writing (see \S~3).

\begin{table}
\begin{center}
\caption{Established showers likely related to asteroids and dual-status objects, order by dates of maximum.\label{tbl:shr}}
\begin{tabular}{lcc}
\hline
Meteor shower & Parent body & Peak date (approx.) \\
\hline
Quadrantids & (196256) 2003 EH$_1$ & Jan. 4 \\
Northern $\delta$ Cancrids & (85182) 1991 AQ & Jan. 16 \\
Southern $\delta$ Cancrids & 2001 YB$_5$ & Jan. 16 \\
Daytime $\kappa$ Aquariids & 2002 EV$_{11}$ & Mar. 20 \\
Daytime April Piscids & (242643) 2005 NZ$_6$ & Apr. 15 \\
$\alpha$ Virginids & 1998 SH$_2$ & Apr. 21 \\
Corvids & (374038) 2004 HW & Jun. 16 \\
Daytime $\beta$ Taurids & 2004 TG$_{10}$ & Jun. 28 \\
$\psi$ Cassiopeiids & (5496) 1973 NA & Jul. 21 \\
$\alpha$ Capricornids & 169P/2002 EX$_{12}$ (NEAT), 2017 MB$_1$ & Jul. 31 \\
$\kappa$ Cygnids & (153311) 2001 MG$_1$, (361861) 2008 ED$_{69}$ & Aug. 13 \\
Northern $\iota$ Aquariids & (455426) 2003 MT$_9$ & Aug. 20 \\
Daytime Sextantids & (155140) 2005 UD & Sep. 30 \\
Northern Taurids & 2004 TG$_{10}$ & Nov. 6 \\
Southern $\chi$ Orionids & 2002 XM$_{35}$, 2010 LU$_{108}$ & Nov. 24 \\
Phoenicids & 289P/2003 WY$_{25}$ (Blanpain) & Dec. 5 \\
Geminids & (3200) Phaethon & Dec. 14 \\
\hline
\end{tabular}
\end{center}
\end{table}

In this review, we will focus on the meteor showers originated from active asteroids and possible dormant comets and discuss how they can help us to understand the evolution of their parent bodies. In \S~2 we discuss the Dissimilarity Criterion and its usage in the identification of parent-shower association. In \S~3 we review the linkages being proposed for established showers as summarized in Table~\ref{tbl:shr}. In \S~4 we discuss how meteor observation can help us to understand comet evolution and highlight some of the recent advances. We conclude this review by a discussion of outstanding problems.

\section{The Dissimilarity Criterion and Its Statistical Significance}

The issue of comet/asteroid-shower association is not an easy one. Most modern search of comet/asteroid-shower association make use of the Dissimilarity Criterion or the $D$ criterion, which was first proposed by \citet{Southworth1963} and has been modified by others \citep[e.g.][]{Drummond1981,Jopek1993,Asher1994,Drummond2000}. A smaller $D$ indicates a higher degree of similarity between two orbits. It is not possible to derive a minimum cut-off of $D$ that corresponds to definite associations, albeit an empirical cut-off of $D\approx0.1$ has been widely used. The issue is further complicated by the fact that the orbits of most showers-of-interest are not precisely known, and that the orbits of meteor showers are also evolving over time.

The original definition of the $D$ criterion given by \citet{Southworth1963} goes as

\begin{equation}
D_{A, B}^2 = \left(q_B - q_A \right)^2 + \left( e_B - e_A \right)^2 + \left( 2\sin{ \frac{I}{2} } \right)^2 + \left[ \left(e_A + e_B \right) \sin{ \frac{\varPi}{2} } \right]^2
\end{equation}

\noindent where
 
\begin{equation}
I = \arccos{ \left[ \cos{i_A} \cos{i_B} + \sin{i_A} \sin{i_B} \cos{\left( \varOmega_A - \varOmega_B \right)} \right]} 
\end{equation}
 
\begin{equation}
\varPi = \omega_A - \omega_B + 2 \arcsin{ \left( \cos{\frac{i_A+i_B}{2}} \sin{\frac{\varOmega_A-\varOmega_B}{2}} \sec{\frac{I}{2}} \right) } 
\end{equation}

\noindent and the subscripts $A$ and $B$ refer to the two orbits being compared. Here $q$ is the perihelion distance in au, $e$ is the eccentricity, $i$ is the inclination, $\Omega$ is the longitude of ascending node, and $\omega$ is the argument of perihelion. The sign of the $\arcsin$ term in the equation for $\varPi$ switches if $|\Omega_A-\Omega_B|>180^\circ$. Most of the later variants to the $D$ criterion similarly rely on the conventional orbital elements.

Since the $D$ criterion only measures the degree of (dis-)similarity of two orbits, it provides limited information on whether the two orbits are likely related. For example, it is common to find likely ``parents'' for ecliptic showers solely based on the $D$ criterion and a simple cutoff at $D=0.1$, since the orbits of most NEOs lie close to the ecliptic plane. To solve this dilemma, we need to calculate the statistical significance of a given $D$: consider the $D$ criterion between the proposed parent-shower pair to be $D_0$, what is the expected number of parent bodies $\langle X \rangle$ that have orbits such that $D<D_0$, where $D$ is the $D$ criterion between the new parent and the shower? 

This topic was first explored by \citet{Wiegert2004} using a debiased NEO population model developed by \citet{Bottke2002}. (Earlier, \citet{Drummond2000} used similar technique to search for groupings of near-Earth asteroids.) More recently, \citet{Ye2016} tested the statistical significances of 32 previously proposed parent-shower pairs with comet-like orbits and found that only 1/4 of them are statistically significant (i.e. $\langle X \rangle \ll 1$). Here we repeat this test\footnote{The script that is used to calculate $\langle X \rangle$ is available at the author's GitHub repository: \url{https://github.com/Yeqzids/d-check}.} to all proposed pairs in Table~\ref{tbl:shr} which includes both cometary and asteroidal showers, using shower orbits derived from contemporary radar and video meteor orbit surveys. $\langle X \rangle$ is computed for NEO population of km-sized objects, since the small masses of objects $\ll1$~km cannot sustain a detectable meteoroid stream \citep{Hughes1989}. For the interest of computing resource and time, we only test 1000 randomly generated NEO populations, therefore our sensitivity of $\langle X \rangle$ only goes down to 0.001. We note, however, that this limit already reaches the $3\sigma$ level which we believe is sufficient to suggest a high confidence linkage.

\begin{table}
\begin{center}
\caption{Statistical significance of the parent-shower linkages in Table~\ref{tbl:shr} assuming that the parent is km-sized. The parent size assumption is valid for most parent bodies with the exception of 289P/Blanpain and 2002 XM$_{35}$, which are $\sim150$~m in diameter assuming a $5\%$ albedo. $\langle X \rangle$ for these two bodies appropriate to their sizes will be $\sim10$ times larger than the values shown in the table. Reference abbreviations are: N64 -- \citet{Nilsson1964}; G75 -- \citet{Gartrell1975}; B08 -- \citet{Brown2008}; B10 -- \citet{Brown2010}; J16 -- \citet{Jenniskens2016}; J16a -- \citet{Jenniskens2016b}; S17 -- \citet{Sato2017}. We note that the numbers for Southern $\chi$ Orionids and Southern $\delta$ Cancrids are uncertain as the orbits of 2001 YB$_5$, 2002 XM$_{35}$ and 2010 LU$_{108}$ are poorly known. \label{tbl:shr1}}
\begin{tabular}{lccc}
\hline
Pair & Reference of parent's orbit & $\langle X \rangle$ for radar orbit & $\langle X \rangle$ for video orbit \\
\hline
(3200) Phaethon --- Geminids & JPL~578 & B10: $0.001$  & J16: $0.001$ \\
2017 MB$_1$ --- $\alpha$ Capricornids & JPL~34 & B08: $0.004$ & J16: $0.004$ \\
(196256) 2003 EH$_1$ --- Quadrantids & JPL~29 & B10: $0.009$ & J16: $0.005$ \\
289P/2003 WY$_{25}$ (Blanpain) --- Phoenicids & JPL~5 & S17: $0.02$  & S17: $0.001$ \\
(155140) 2005 UD --- Daytime Sextantids & JPL~66 & B10: $0.1$ & J16: $0.05$ \\
(374038) 2004 HW --- Corvids & JPL~60 & -- & J16: $0.1$ \\
1998 SH$_2$ --- $\alpha$ Virginids & JPL~121 & -- & J16a: $0.2$ \\
2004 TG$_{10}$ --- Daytime $\beta$ Taurids & JPL~25 & B08: $0.2$  & -- \\
2004 TG$_{10}$ --- Northern Taurids & JPL~25 & B08: $0.3$  & J16: $0.1$ \\
169P/2002 EX$_{12}$ (NEAT) --- $\alpha$ Capricornids & JPL~121 & B08: $0.3$  & J16: $0.3$ \\
(85182) 1991 AQ --- Northern $\delta$ Cancrids & JPL~83 & -- & J16: $0.3$ \\
(242643) 2005 NZ$_6$ --- Daytime April Piscids & JPL~73 & B10: $0.5$ & -- \\
(153311) 2001 MG$_1$ --- $\kappa$ Cygnids & JPL~63 & -- & J16: $0.8$ \\
(5496) 1973 NA --- $\psi$ Cassiopeiids & JPL~52 & B08: $1$ & J16: $0.5$ \\
(361861) 2008 ED$_{69}$ --- $\kappa$ Cygnids & JPL~36 & -- & J16: $2$ \\
2002 EV$_{11}$ --- Daytime $\kappa$ Aquariids & JPL~20 & G75: $2$ & -- \\
2010 LU$_{108}$ --- Southern $\chi$ Orionids & JPL~12 & N64: $2$ & J16: $0.6$ \\
2002 XM$_{35}$ --- Southern $\chi$ Orionids & JPL~7 & N64: $4$ & J16: $0.5$ \\
2001 YB$_5$ --- Southern $\delta$ Cancrids & JPL~6 & N64: $9$ & J16: $0.4$ \\
(455426) 2003 MT$_9$ --- Northern $\iota$ Aquariids & JPL~38 & B10: $18$ & J16: $0.05$ \\
\hline
\end{tabular}
\end{center}
\end{table}

As shown in Table~\ref{tbl:shr1}, our calculation confirmed some of the well-known linkages such as the (3200) Phaethon --- Geminids pair and the (196256) 2003 EH$_1$ --- Quadrantids pair, while some of the linkages such as the (455426) 2003 MT$_9$ --- Northern $\iota$ Aquariids are found to be statistically unlikely. While the results derived from radar and video orbits agree in most cases, there are a few cases where radar result and video result deviates from each other, such as 2002 XM$_{35}$ --- Southern $\chi$ Orionids, 2001 YB$_5$ --- Southern $\delta$ Cancrids and (455426) 2003 MT$_9$ --- Northern $\iota$ Aquariids. Radar orbits of the first two showers are based on very small statistics which could explain the deviation from the video orbits. However, the shower for the last case, Northern $\iota$ Aquariids, is a well-observed shower. It would be interesting to investigate the discrepancy between the the radar and the video orbits of Northern $\iota$ Aquariids though it is beyond the scope of this review.

Before we discuss high confidence linkages, which we will do in the next section, let us reflect on the complication arisen from the dynamical evolution of the meteoroid stream, a process that dissociate parent-shower linkage over time. To understand how the dynamical evolution of meteoroid streams affects $\langle X \rangle$, we conduct a simple experiment on four objects: (3200) Phaethon, (196256) 2003 EH$_1$, 2004 TG$_{10}$, and 209P/LINEAR, the latter of which is the parent of the Camelopardalid meteor shower on the IAU working list \citep{Jenniskens2006,Ye2014}. We select these four objects as they are well known as shower parents and cover a relatively wide orbital and $\langle X \rangle$ spaces. For each object, one 1~mm particle representing the median of the associated meteoroid stream is released at zero speed with respect to the parent. The choice of 1~mm reflects the typical sizes of meteoroids detectable by most conventional techniques \citep{Ceplecha1998} and is meant to simplify our discussion, though we note that $\langle X \rangle$ (and also $D$) is somewhat dependent on the size distribution of meteoroids. The parent and the particles are integrated forward for $10^4$~yr using a tailored Mercury6 package \citep{Chambers1999,Ye2016a}, considering gravitational perturbation from major planets (with the Earth-Moon system represented by a single perturber), radiation pressure and Poynting-Robertson effect. Parents and all particles are considered massless and do not interact with each other. The choice of an integration duration of $10^4$~yr is made considering the collisional lifetime of millimeter-sized meteoroids \citep{Grun1985}. Orbits of the parents and the particles are recorded every 100~yr with their $\langle X \rangle$ values being calculated following the aforementioned procedure.

\begin{figure}
\label{fig:x}
\includegraphics[width=\textwidth]{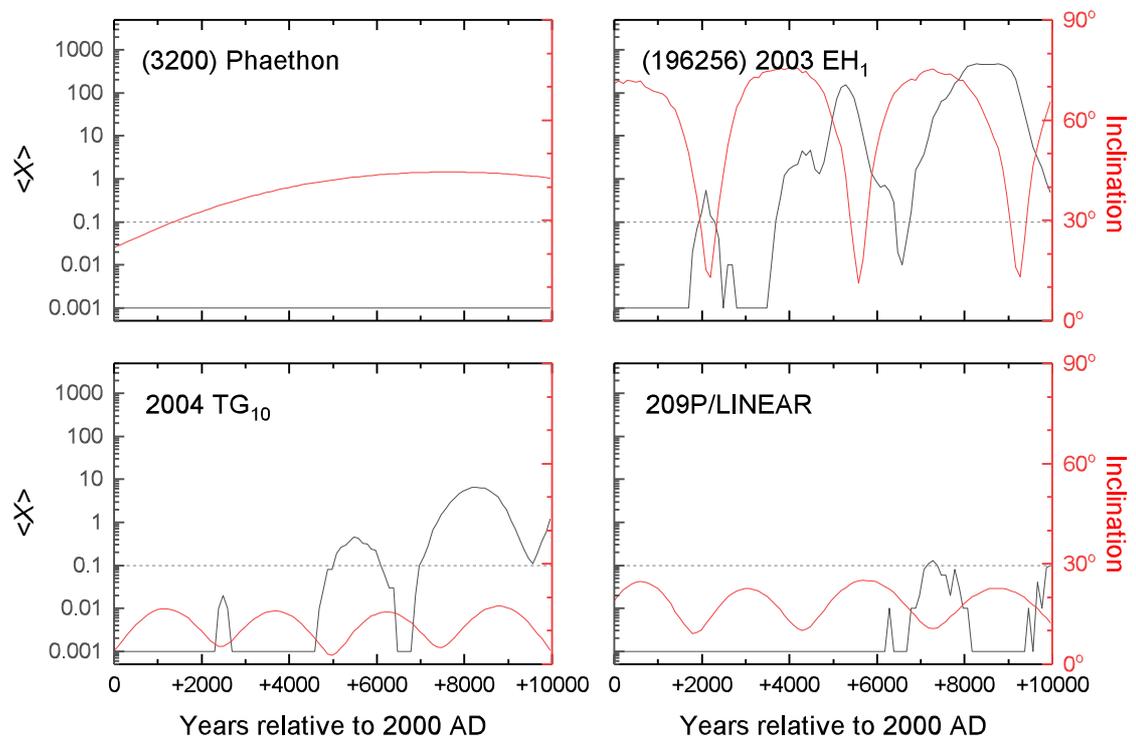}
\caption{Evolution of $\langle X \rangle$ (black line) and inclination (red line) of 1~mm particles released by (3200) Phaethon, (196256) 2003 EH$_1$, 2004 TG$_{10}$, and 209P/LINEAR. Note that the metrics of the parents are not explicitly plotted. The critical cutoff of $\langle X \rangle=0.1$ is depicted in a dashed line.}
\end{figure}

In Figure~\ref{fig:x} we show the evolution of $\langle X \rangle$ over $10^4$~yr for each of the four targets being tested. Again, our sensitivity of $\langle X \rangle$ only goes down to 0.001 as we only test 1000 synthetic NEO populations. We find that meteoroid stream generated by (3200) Phaethon stay in a very stable orbit, allowing parent-shower association to be made beyond a timescale of $10^4$~yr. This is likely due to the fact that the orbit of Phaethon prevents it from close approach with large major planets like Jupiter. The other three objects make regular approaches to Jupiter and therefore their streams are less stable. It only takes a few $10^3$~yr for streams produced by (196256) 2003 EH$_1$ and 2004 TG$_{10}$ to become statistically detached from their parents, while the stream by 209P/LINEAR, which is known to be residing in a stable resonance point \citep{Fernandez2015,Ye2016a}, take over $10^4$~yr to become decoherent with its parent. Nevertheless, our experiment suggests that parent-shower pairs similar to the cases of (3200) Phaethon, (196256) 2003 EH$_1$, 2004 TG$_{10}$, and 209P/LINEAR should remain statistically identifiable for at least a few $10^3$~yr, a timescale consistent to the age of typical meteoroid streams \citep{Pauls2005}, though very massive and Jupiter-approaching streams that can survive over $10^4$~yr could indeed be detached from their parents, making parent-shower association very difficult.

\section{High Confidence Parent-Shower Linkages}

Here we review the high confidence parent-shower linkages identified in Table~\ref{tbl:shr1} order by their statistical significances.

\paragraph{(3200) Phaethon --- Geminids}
Identified in 1983 and being associated to one of the strongest annual meteor showers, the Phaethon --- Geminids pair is the earliest identified and perhaps the best known asteroid-shower pair. We find the likelihood of chance alignment to be 1 in 1000 for both radar- and video-derived orbits, suggesting that the Phaethon --- Geminids pair is likely to be genuine as expected. The formation mechanism of the Geminids is still under debate, with asteroidal collision \citep{Hunt1986}, cometary sublimation \citep{Gustafson1989}, and thermal evolution \citep{Kasuga2009} having been proposed as likely driver. It has been recently found that Phaethon does currently show some weak activity at its extreme perihelion of $q=0.14$~au \citep{Jewitt2010}, albeit the dust production level is too small to explain the formation of the Geminid meteoroid stream.

\paragraph{2017 MB$_1$ --- $\alpha$ Capricornids}
The $\alpha$ Capricornid meteor shower was originally associated to 169P/2002 EX$_{12}$ (NEAT) \citep{Wiegert2004,Jenniskens2010,Kasuga2010} with a likelihood of chance alignment to be 1 in 3. However, a recently-found asteroid, 2017 MB$_1$, appears to be a much better parent candidate \citep{Wiegert2017}, with a 1 in 250 chance to be coincidence. 169P/NEAT is a weakly active comet while 2017 MB$_1$ has not been found to be currently active. Numerical simulation shows that, assuming 169P/NEAT and $\alpha$ Capricornids is physically related, a major disruption took place on the comet about 4500--5000 years ago that lead to the formation of the meteoroid stream. Dust released at an earlier or later epoch would not reach the Earth at the right time to be currently observable. An interesting possibility is that 169P/NEAT, 2017 MB$_1$ and $\alpha$ Capricornids all belong to a larger progenitor that underwent a large fragmentation $\sim5000$~years ago, though a critical examination is needed.

\paragraph{(196256) 2003 EH$_1$ --- Quadrantids}
The Quadrantids is the second identified asteroidal shower after the Geminids \citep{Jenniskens2004,Williams2004}. Our calculated likelihood of chance alignment is 1 in 100--200, much higher than the 1 in 2 million rate given by \citet{Jenniskens2008}, but still within reasonable range that suggests a likely linkage. (196256) 2003 EH$_1$ is about 2~km in size and has an orbit comparable to most short-period comets, yet none of the attempts to search for cometary activity have been successful \citep{Kasuga2015}. The young dynamical age of the Quadrantid meteoroid stream, which is 200--500~yr \citep{Wiegert2004a,Abedin2015}, implies that (196256) 2003 EH$_1$ (or its true parent) must have been active within the recent a few hundred years. More broadly, the (196256) 2003 EH$_1$ --- Quadrantids pair joins several other notable comet/asteroid-shower pairs to become what is known as the Machholz complex, named after comet 96P/Machholz. It is believed that this renown complex is originated from cascading fragmentation of 96P/Machholz over the previous $\sim10^4$~yr \citep{Abedin2018}.

\paragraph{289P/2003 WY$_{25}$ (Blanpain) --- Phoenicids}
Independently identified by \citet{Micheli2005} and \citet{Jenniskens2005a}, this linkage convincingly resolves the mysteries over the long-lost comet D/1819 W1 (Blanpain) and origin of the Phoenicid meteor shower. It is hypothesized that the progenitor of 289P/Blanpain experienced a series of fragmentation events in 1817--1819, which produced a large amount of dust that helped its discovery, as well as at least one smaller remnant that is currently known as 289P/Blanpain. Dust released in 1819 approached the Earth in 1956 and 2014, with heighten meteor activity confirmed by meteor observations \citep{Watanabe2005, Sato2010, Sato2017}. The likelihood of chance alignment is calculated to be 1 in 50--1000, supporting the idea that 289P/Blanpain and the Phoenicids are related. It has been found that 289P/Blanpain is still weakly active, at a level that is too low to replenish the Phoenicid stream \citep{Jewitt2006}.

\paragraph{(155140) 2005 UD --- Daytime Sextantids}
The (155140) 2005 UD --- Daytime Sextantids pair joins the Phaethon --- Geminids and asteroid (225416) 1999 YC, forming the so-called Phaethon-Geminids Complex (PGC) \citep{Ohtsuka2006, Ryabova2008, Kasuga2009}. It has been proposed that these bodies and streams were formed as a result of thermal disintegration of a much larger progenitor \cite{Kasuga2009}. Our calculation shows a likelihood of chance alignment for the (155140) 2005 UD --- Daytime Sextantids pair to be 1 in 10--20. Study of the Daytime Sextantids is scarce despite the fact that the shower is quite strong and has been observed by both radar and video techniques \citep{Brown2010, Jenniskens2016}.

\paragraph{(374038) 2004 HW --- Corvids}
The Corvids was reported only by \citet{Hoffmeister1948} based on visual data before the recent confirmation by the Cameras for Allsky Meteor Surveillance (CAMS) network based on 12 meteors \citep{Jenniskens2016}. Its southerly radiant, combining with a very low geocentric encounter speed ($\sim9$~km/s), making detection and confirmation difficult. Our calculation shows a likelihood of chance alignment to be 1 in 10. More orbit measurement is encouraged in order to further verify this linkage.

\paragraph{2004 TG$_{10}$ --- Northern Taurids}
As a member of the Taurid Complex, the Northern Taurids is typically being associated to comet 2P/Encke, though more than 10 asteroids have been proposed to be members of this complex, including 2004 TG$_{10}$ \citep{Porubvcan2006,Babadzhanov2008,Olech2017,Spurny2017}. (The Daytime $\beta$ Taurid meteor shower in Table~\ref{tbl:shr1} is also a member of the Taurid Complex.) Our calculation shows a moderate chance for Northern Taurids --- 2004 TG$_{10}$ to be a chance alignment (1 in 10 to 1 in 3). However, this number should be taken cautiously due to the complicated dynamical history of the Taurid Complex.

\section{Meteor Observation as a Tool to Understand Comet Evolution}

Since meteors are related to previous activities of the parent, they provide some information about the history of the parent. These information can be very valuable if they are from times that the parent had not yet been discovered. However, we also need to recognize that meteor observation can only provide a very skewed picture of what has happened to the parent: only the dust that are presently intercepting Earth's orbit can be detected as meteors.

The case of the now-defunct comet 3D/Biela is the earliest and the perhaps the best example of what meteor observation can do to help understand comet evolution. Biela's Comet was initially discovered by Jacques Leibax Montaigne in 1772 and was named after its orbit computer Wilhelm von Biela \citep{Kronk1999,Kronk2003}. The comet was found to have split during its 1846 perihelion and was lost after its 1852 perihelion. However, spectacular meteor storms radiating from the constellation of Andromeda was observed in 1872 and 1885, with orbits consistent with Biela's Comet, suggesting a complete disintegration of the comet \citep{Olivier1925}. Recent observation of the Andromedids, coupled with dynamical simulation, suggests that Biela's Comet had been active for at least $\sim200$~yr before its final disintegration \citep{Wiegert2013}. A comprehensive analysis of the remaining mass of dust in the Andromedid meteoroid stream suggests that some larger fragments of Biela's Comet may have survived the disintegration and is now hiding as a dormant comet, though such fragments (if exist) are yet to be found \citep{Jenniskens2007}.

The case of 289P/Blanpain, introduced in the previous section, is another example. Compared to the case of 3D/Biela, a large remnant that survived the fragmentation has actually been recovered and still exhibits some very low activity \citep{Jewitt2006}. A comprehensive analysis that make use of the available telescopic and meteor data, coupled with dynamical simulation, should provide a better picture of the fragmentation process. Events like 3D/Biela and 289P/Blanpain are unique as they allow us to directly sample the dust deposited by the parent, providing useful analogues to events like the fragmentation of 332P/Ikeya-Murakami \citep{Ishiguro2014,Jewitt2016,Kleyna2016,Hui2017}, that can only be studied by telescopic observations as the parents do not approach the Earth.

Meteor observation can also reveal historic episodic ejection of now-dormant parents. One of the examples is the 2006 outburst of June $\alpha$ Virginids, likely associated to asteroid (139359) 2001 ME$_1$ with a chance alignment rate of 1 in 100 \citep{Ye2016}. This event can be considered as an analogue to the possible transient ejection of 107P/(4015) Wilson-Harrington in 1949 \citep{Fernandez1997}.

Meteor surveys also find a number of \textit{orphan} showers that cannot be associated to any known asteroids or comets. We cannot exclude the possibility that the parent bodies are yet to be found, but given that our knowledge of km-sized near-Earth objects are now $>90\%$ complete \citep{Jedicke2015}, it is likely that at least some of the short-period showers are originated from catastrophic disintegration of comets or asteroids. It has been suggested that near-Sun asteroids could disrupt due to intense thermal effects, leaving behind orphan streams \citep{Granvik2016}. Examination of telescopic survey data also suggest that comet disruptions may be common \citep{Ye2017}. In theory, meteor data could provide an independent constraint of the number of near-Earth asteroids or comets that have recently disintegrated.

\section{Future Work}

A lot of exciting advancements have been made since the review of \citet{Jenniskens2008}. Four large video surveys have since been built or greatly expanded, providing almost 1~million new video meteoroid orbits compared to less than 80,000 ten years ago \citep{Jenniskens2017}. Video networks specifically aiming at meteorite recovery have been built or greatly expanded \citep{Bland2012,Madiedo2014i,Colas2016a}, enhancing our chances of recovering meteorites from slow showers such as the Geminids and the Taurids \citep{Madiedo2013a,Madiedo2014b}. The Canadian Meteor Orbit Radar (CMOR) has been upgraded and has measured 14~million meteoroid orbit since 2002 \citep{Ye2013,Ye2016}. The Southern Argentina Agile Meteor Radar (SAAMER) has been set up to patrol the southern sky \citep{Janches2013,Janches2014,Janches2015}. Various radar systems occasionally conduct meteor observations \citep{Janches2008,Kero2012,Younger2015}. The rapid increase of meteor orbit data is particularly encouraging for the studies of weakly active showers likely originated from active asteroids and dormant comets.

One major problem that is yet to be convincingly solved is the identification of weakly active showers. Traditional practice of manually identifying radiant ``clusters'' is difficult to cope with the large data rate of modern meteor surveys. Methods that are widely used by modern surveys include wavelet transformation \citep{Brown2008,Brown2010} and clustering linkage \citep{Rudawska2015,Jenniskens2016}. However, both techniques still rely on a number of unconstrained free parameters such as radiant sizes and velocity spreads, therefore unique identification is difficult for very weak showers close to the background. New techniques, such as a variable critical $D$ criterion \citep{Moorhead2016} and comparison with synthetic orbits \citep{Vida2017}, are being explored to overcome this problem.

Contemporary NEO surveys have found a number of intriguing objects like (3200) Phaethon, (196256) 2003 EH$_1$ and others, many with (or likely to have) associated meteoroid streams. Besides a few notable ones, most of these complexes are poorly understood. For example, studies remain scarce for the members in the Phaethon-Geminid Complex besides Phaethon --- Geminids itself; most objects and meteor showers in various sungrazing and sunskirting families remain to be characterized \citep{Sekhar2014}.

\section*{Acknowledgment}

I thank David Asher and an anonymous reviewer for their comments that help improve this manuscript. This work is supported by the GROWTH project funded by the National Science Foundation under Grant No. 1545949.



%% References


\bibliographystyle{elsart-harv}
\bibliography{ms}






% \end{thebibliography}

\end{document}